%% file: main.tex
\documentclass[12pt]{iopart}

\usepackage{natbib}
  \expandafter\let\csname equation*\endcsname\relax
  \expandafter\let\csname endequation*\endcsname\relax
\usepackage{amsmath}
\usepackage{graphics}
\usepackage{graphicx}
\usepackage{caption}
\usepackage{color}
\usepackage{soul}
\usepackage{subfig}
\usepackage{booktabs}
\usepackage{makecell}
\usepackage[symbol]{footmisc}
\usepackage[ruled,vlined]{algorithm2e}

\SetKwInput{kwInput}{Input}
\SetKwInput{kwOutput}{Output}
\SetKwIF{If}{ElseIf}{Else}{if}{}{else if}{else}{}%
\SetKwFor{For}{foreach}{}{}%
\SetKwRepeat{Do}{do}{for}
\SetKw{KwGoTo}{go to}

\setstcolor{red}
\bibpunct{(}{)}{;}{a}{,}{,}

\usepackage{hyperref}
\hypersetup{
    colorlinks=true,
    linkcolor=blue,
    filecolor=magenta,      
    urlcolor=cyan,
}

\begin{document}


\title[Identification of 27 abnormalities from multi-lead ECG signals]{Identification of 27 abnormalities from multi-lead ECG signals: An ensembled SE\_ResNet framework with Sign Loss function}

\author{ 
Zhaowei Zhu$^{1}$\footnote{These two authors are jointly first authors.\label{first_author}},%
Xiang Lan$^{2}$\text{\ref{first_author}},
Tingting Zhao$^{1}$,
Yangming Guo$^{1}$,
Pipin Kojodjojo$^{3}$,
Zhuoyang Xu$^{1}$,
Zhuo Liu$^{1}$,
Siqi Liu$^{4}$,
Han Wang$^{2}$,
Xingzhi Sun$^{1}$\footnote{These two authors are jointly corresponding authors.\label{cor_author}},
Mengling Feng$^{2,5}$\text{\ref{cor_author}}
\\
\small{
$^1$ Ping An Technology, Beijing, China\\
$^2$ Saw Swee Hock School of Public Health, National University Health System, National University of Singapore,  Singapore\\
$^3$ Yong Loo Lin School of Medicine, National University Health System, National University of Singapore,  Singapore\\
$^4$ NUS Graduate School for Integrative Sciences \& Engineering, Singapore\\
$^5$ Institute of Data Science, National University of Singapore,  Singapore\\
}}

\ead{zhuzhaowei262@pingan.com.cn, ephlanx@nus.edu.sg}

\begin{abstract}

\input{0_abstract}

\end{abstract}

\maketitle

\section{Introduction}
\input{1_intro}

\section{Methods}
\input{2_methods}

\section{Results}
\input{3_results}

\section{Discussion}
\input{4_discussion}

\section{Conclusions}
\input{5_conclusions}

\section*{Acknowledgments} 
\input{6_acknowledgments}


\def\newblock{\hskip .11em plus .33em minus .07em}

\bibliographystyle{dcu}
\bibliography{refs}

\end{document}

%% file: 0_abstract.tex
{\it Objective}:
Cardiovascular disease is a major threat to health and one of the primary causes of death globally. The 12-lead ECG is a cheap and commonly accessible tool to identify cardiac abnormalities. Early and accurate diagnosis will allow early treatment and intervention to prevent severe complications of cardiovascular disease. In the PhysioNet/Computing in Cardiology Challenge 2020, our objective is to develop an algorithm that automatically identifies 27 ECG abnormalities from 12-lead ECG recordings.

{\it Approach}:
Firstly, a series of pre-processing methods were proposed and applied on various data sources in order to mitigate the problem of data divergence. Secondly, we ensembled two SE\_ResNet models and one rule-based model to enhance the performance of various ECG abnormalities’ classification. Thirdly, we introduce a Sign Loss to tackle the problem of class imbalance, and thus improve the model’s generalizability. 

{\it Main Results}:
Our proposed approach achieved a challenge validation score of 0.682, and a full test score of 0.514, placed us 3rd out of 40 in the official ranking.

{\it Significance}:
We proposed an accurate and robust predictive framework that combines deep neural networks and clinical knowledge to automatically classify multiple ECG abnormalities. Our framework is able to identify 27 ECG abnormalities from 12-lead ECG data regardless of discrepancies in data sources and the imbalance of data labelling. We trained and validated our framework on 6 datasets from various countries.

%% file: 1_intro.tex
Cardiovascular disease is one of the primary causes of death globally, resulting in estimated 16.7 million deaths  each year, according to the World Health Organization \cite{GAZIANO201072}.  Early and accurate diagnosis of ECG abnormalities can prevent serious complications such as sudden cardiac death  and  improve treatment outcomes \cite{1991}.

The 12-lead ECG is a cheap, widely available tool to screen for heart disease screening \cite{kligfield2007recommendations}. However, interpretation of the ECG requires  experienced clinicians to carefully examine and  recognize pathological inter-beat and intra-beat patterns. This process is time-consuming and subject to inter-observer variability \cite{bickerton_pooler_2019}. Hence, an accurate algorithm for automated ECG pattern classification is  highly desirable.

Some earlier works have reported automated analysis of ECG \cite{martinez2004wavelet, minami1999real, mahmoodabadi2005ecg, alexakis2003feature}. These approaches are mainly based on frequency domain features, time-frequency analysis, and signal transformations (i.e., Wavelet transform and Fourier transform). However, such techniques are not capable to capture complex features of the ECG signal.

More recently, a number of works have demonstrated the ability of non-linear machine learning techniques in the field of ECG analysis. \cite{VAFAIE2014291} proposed a classifier to predict heart diseases, in which a fuzzy classifier was constructed with a genetic algorithm to improve prediction accuracy. \cite{chen2018classification} designed a gradient boosting algorithm to detect atrial fibrillation. Piece-wise linear splines were used to select features. The performance of those approaches is limited by the choices of input features to the models. The input features are typically selected by sophistic, which requires time and expert knowledge to verify the feasibility. 

In recent years, deep learning and neural networks, especially the Convolutional Neural Networks (CNNs) \cite{lecun1995convolutional}, have gained promising results in many areas such as computer vision \cite{krizhevsky2017imagenet} and natural language processing \cite{devlin2018bert}. For ECG abnormalities detection based on ECG signal, \cite{xiong2018ecg} developed a 21-layer 1D convolutional recurrent neural network to detect atrial fibrillation, which was trained on single-lead ECG data. \cite{sodmann2018convolutional} proposed a CNN model to improve detection performance by annotating QRS complexes, P waves, T waves, noise, and inter-beat of ECG segments.  \cite{warrick2018ensembling} designed an ensemble deep learning model for automatic classification of cardiac arrhythmias based on single lead ECGs, which fused the decisions of ten classifiers and outperformed the single deep classifier.

Most of the previous works focus on one or at most 9 ECG abnormalities's classification \cite{2019A,0Multi, 2019Arrhythmia,2019Automatic}. Therefore, we aim to develop a robust model that is able to generalize to 27 different types of ECG abnormalities. In addition, most of the existing works only deal with single-lead ECG signals \cite{0Comparing,0Detection,2017Detection, warrick2018ensembling}, while in the clinical practice, 12-lead ECGs are more commonly used for abnormality detection and diagnosis. Hence, we developed a model that analyses all 12-lead ECG signals to capture the full picture of potential abnormalities.  

The novel contributions of this work include:

1) We aim to develop a robust model that is able to identify 27 types of ECG abnormalities by analysing 12-lead ECG signals from multiple datasets.

2) From our statistical analysis of the dataset, we found a marked class imbalance of the dataset, where the number of certain types of ECG abnormalities’ is 40 times more than the others. Consequently, the model will likely perform poorly in the minority class if we don’t handle the class imbalance properly. Therefore, we propose to use Sign Loss function to mitigate the negative effects caused by class imbalance.

3) Unlike the aforementioned works that were developed on a small dataset from a single data source. We have trained and validated the performance of our model over 6 different datasets across the world. In addition, we proposed a series of multi-source ECG data pre-processing method to improve the generalizability of our proposed model.

%% file: 2_methods.tex
\begin{figure}[t]
\begin{centering}
\includegraphics[scale=0.26]{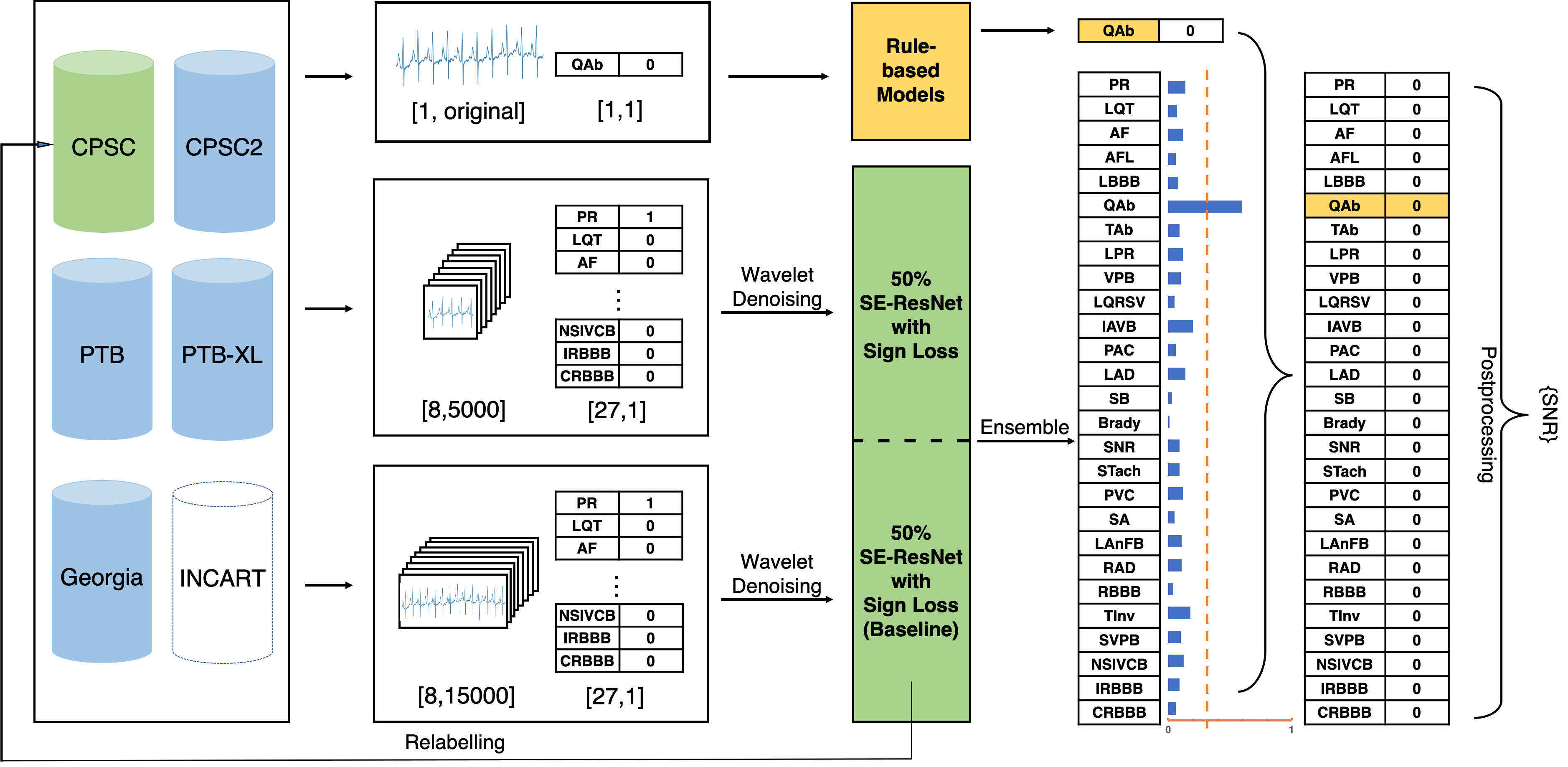}
\caption{Illustration of the framework design. To enable the model train over 6 different dataset, a series of data preprocessing method was proposed and applied in the framework. We ensembled two SE\_ResNet and one rule-based model to improve the model's performance and efficiency. Both SE\_ResNets are using a special multi-label Sign Loss function to mitigate the class imbalance problem.}
\label{fig:1}
\end{centering}
\end{figure}

The overall framework design is shown in Figure \ref{fig:1} and our methods will be elaborated below. 

\subsection{Datasets}


\textbf{Training dataset}. The public challenge data consist of 43,101 12-lead ECG signals from 6 different datasets, including 2 datasets from China, 2 datasets from German, 1 datasets from Russia. They are China Physiological Signal Challenge 2018 (CPSC), China Physiological Signal Challenge 2018 Extra (CPSC2), Physikalisch Technische Bundesanstalt Diagnostic ECG Database (PTB), Physikalisch Technische Bundesanstalt Electrocardiography Database (PTB-XL), Georgia 12-lead ECG Challenge Database (Georgia), and St.Petersburg Institute of Cardiological Technics 12-lead Arrhythmia Database (INCART). The sampling frequency of the signals varies from 257 Hz to 1,000 Hz, and the length of the signals varies from 6 seconds to 30 minutes. There are 111 labelled abnormalities, where multiple labels may correspond to the same type of abnormality. Therefore, 27 types of abnormalities are included in the final scoring metrics.  From these data, we created our offline training set and validation set. Table 1 shows basic information of the 6 datasets.


\textbf{Offline validation dataset}. After we processed original training data as described in 2.4, we randomly split 80\% as the training set (30,172 samples) and 20\% as our offline validation set(7,544 samples). To further validate the generalizability of our model, an external dataset from the Hefei Hi-tech Cup ECG Intelligent Competition \cite{hefei} (Hefei dataset in short) was applied as external validation. Hefei dataset consists of 40,000 records of 8-lead ECG signals with a sampling frequency of 500 Hz and length of 10 seconds. Out of all the records, 6,500 records with labels in the 27 types of ECG abnormalities that we focused were randomly selected and formed an external validation set. 


\begin{table}[t]
\centering
\begin{tabular}{|l|l|l|l|l|l|}
\hline
Databases & Recordings & Labels  & Extra labels & Extra Recordings \\ \hline
CPSC & 6,877 & 9  & - & - \\ \hline
CPSC2 & 3,453 & 72  & - & - \\ \hline
INCART & 74 & 37  & 11 & 53 \\ \hline
PTB & 516 & 17  & 7 & 46 \\ \hline
PTB-XL & 21,837 & 50  & - & - \\ \hline
Georgia & 10,344 & 67  & - & - \\ \hline
\end{tabular}
\caption{\label{table1} Basic information of six different datasets that forms the challenge dataset.}
\end{table}


\textbf{Online test dataset}. The entire online test data contains 11,630 12-lead ECG recordings that were not represented in the training data. The test data were drawn from three databases shown below.

Test Database 1: A total of 1,463 ECGs from Southeast University, China, including the data from the China Physiological Signal Challenge 2018.

Test Database 2: A total of 5,167 ECGs from the Georgia 12-Lead ECG Challenge Database, Emory University, Atlanta, Georgia, USA.

Test Database 3: A total of 10,000 ECGs from an unspecified US institution comparable to Test Database 2, matched for demographics and prevalence of classes.

In addition, an official validation set was used by the challenge, the source and size of the official validation set have not published by the challenge organizer.

\subsection{Multi-source Data Pre-processing}
The raw data were sampled from different sources, which varies in sampling rate, signal amplitude, noise level, etc. To better prepare the data for model training, we adopted the following data pre-processing techniques.

\textbf{Processing original data.}
INCART dataset was excluded from our training data since it has only 74 30-minutes records with a sampling frequency of 257 Hz and is significantly different from other datasets. All data without a label in the 27 scored classes were excluded as well. PTB dataset was downsampled from 1,000 Hz to 500 Hz to make the sampling frequency of all training data consistent. Since lead III, aVR, aVL, and aVF are linearly dependent on other leads and can be calculated based on Einthoven’s Law \cite{kligfield2002centennial} and Goldberger's equations \cite{goldberger_goldberger_shvilkin_2018}, these 4 leads were also excluded.

\textbf{Truncating \& padding.} 
For the baseline model, all input signals were fixed at 30 seconds in length. This was done by truncating the part exceeding the first 30 seconds for longer signals and padding the shorter signals with zero. For the other ensembled model, the input length was fixed at 10 seconds with the same preprocessing method.

\textbf{Wavelet denoising.} To reduce the noise in ECG signals, biorthogonal wavelet transformation was applied. The numbers of vanishing moments for the decomposition and reconstruction filters were 2 and 6 respectively. The level of refinement was set to be 8.

\textbf{Relabelling CPSC data.}
CPSC dataset was relabelled due to the fact that the original labels cover only 9 classes and the distribution of the labels is significantly different from other datasets. A baseline model was first trained on the original training set and then used for inferring pseudo labels on the CPSC dataset. For each ECG signal in the CPSC dataset, inferred pseudo labels were added as new labels if 1) the inference output probability was higher than 0.8, 2) the labels that were not in the original 9 labels and 3) the labels were in the 27 officially scored labels. To check the validity of our relabelling strategy, a clinician reviewed 11 records out of all the relabelled data with inference output probability higher than 0.95. The feedback that most of the new labels were valid testified that the CPSC dataset has missing labels and our pseudo labels were valid.

\subsection{Multi-label Classification With 12-leads ECG}

\textbf{SE\_ResNet.} One important feature of 12-lead ECG signal is the information contained differs in different leads, due to the difference of signal voltage intensity and variation in amplitude Different ECG abnormalities may be clearer in certain leads.  An equal importance of different leads could cause information losses, which lead to misdiagnosis.

\begin{figure}[t]
\begin{centering}
\includegraphics[scale=0.45]{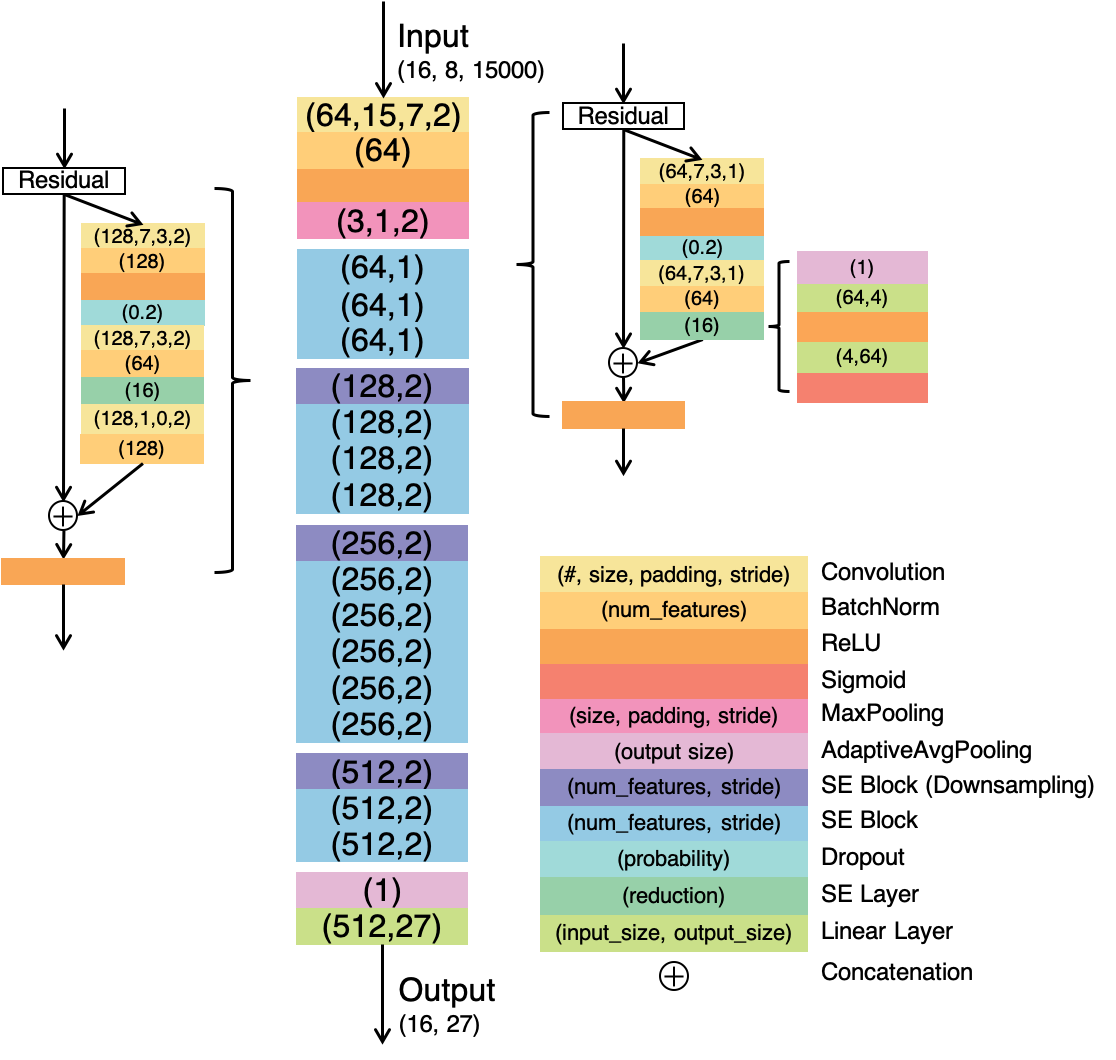}
\caption{Architecture of the SE\_ResNet model.}
\label{fig:3}
\end{centering}
\end{figure}

Therefore, we use SE\_ResNet \cite{SEResNet} as our primary model capture the distinctive information in each of the multi-leads ECG signal. We integrate a Squeeze-and-excitation (Se) block into the ResNet \cite{ResNet} structure. structure. The squeeze operation compresses the global spatial information and produces an embedding of the global distribution of feature responses for each channel, thus all the layers of the network are able to use the information from the global receptive field. The excitation operation takes the embedding as input and captures the channel-wise dependencies, then produces weights for each channel. Consequently, these weights are applied to previously learned feature maps and realize the feature recalibration. 

In this way, some leads could be given higher weights, leading to a better prediction performance for multi-lead ECG classification. Our baseline model was a SE\_ResNet model with an input length of 30 seconds. To minimize the effect of padding on the shorter signals, another SE\_ResNet model was trained with the input length of 10 seconds and ensembled with the baseline model. The structure of our SE\_ResNet model is shown in Figure \ref{fig:3}.

\textbf{Rule-based model.}
The baseline model did not perform well on certain classes, which could be guided by clinical rules. One of such classes is bradycardia, which indicates the heart rate is slower than 1 beat per second. In another word, the R-R intervals between consecutive heartbeats are always longer than 1 second. To detect the R-R intervals, Pan \& Tompkins algorithm \cite{pan32real} was used to detect the R-peaks on lead I. Then, R-R intervals could be easily calculated from the positions of consecutive R-peaks. The pseudocode of the rule-based model for bradycardia is shown in Algorithm 1. 

\begin{algorithm}
\SetAlgoNoLine
\kwInput{List of R-R intervals}
\kwOutput{Classification of bradycardia}

 brady\_beats = 0\;
 \For{R-R interval}{
 \If{1s $\leq$ length of interval $\leq$ 1.6s}
  {brady\_beats += 1}
 }
 \eIf{brady\_beats / \# of R-R intervals $\geq$ 0.5}
 {\Return True}{\Return False}
 \caption{Rule-based bradycardia classifier}
\end{algorithm}

\begin{algorithm}
\kwInput{Prediction from ensembled model\newline Prediction from rule-based model}
\kwOutput{Final classification of bradycardia}
\SetAlgoNoLine
 \If{Prediction from rule-based model is False}
  {\Return False}
 \Else{\Return Prediction from ensembled model}
 \caption{Final bradycardia prediction}
\end{algorithm}

A high recall and low precision were observed when we only applied the rule-based model to classify bradycardia. It could be due to the low quality of labels in the datasets. Therefore, the rule-based model was only applied when the prediction of bradycardia is negative from the ensembled model. The pseudocode for the final bradycardia prediction is shown in Algorithm 2.

\textbf{Model Ensemble.}
Fusion is a common and effective method to improve generalizability. The idea of fusion in this paper is to combine two models that receive different length input signals. Different data lengths could provide the model with multiple views of data. The two lengths selected in this model are 10 seconds and 30 seconds (corresponding to the data length of 5,000 and 15,000). Signals that were predicted to be negative for all classes were revised to be positive for the default normal class, sinus rhythm (SNR).

\subsection{Class Imbalance}

\begin{figure}[t]
\begin{centering}
\includegraphics[width=\textwidth]{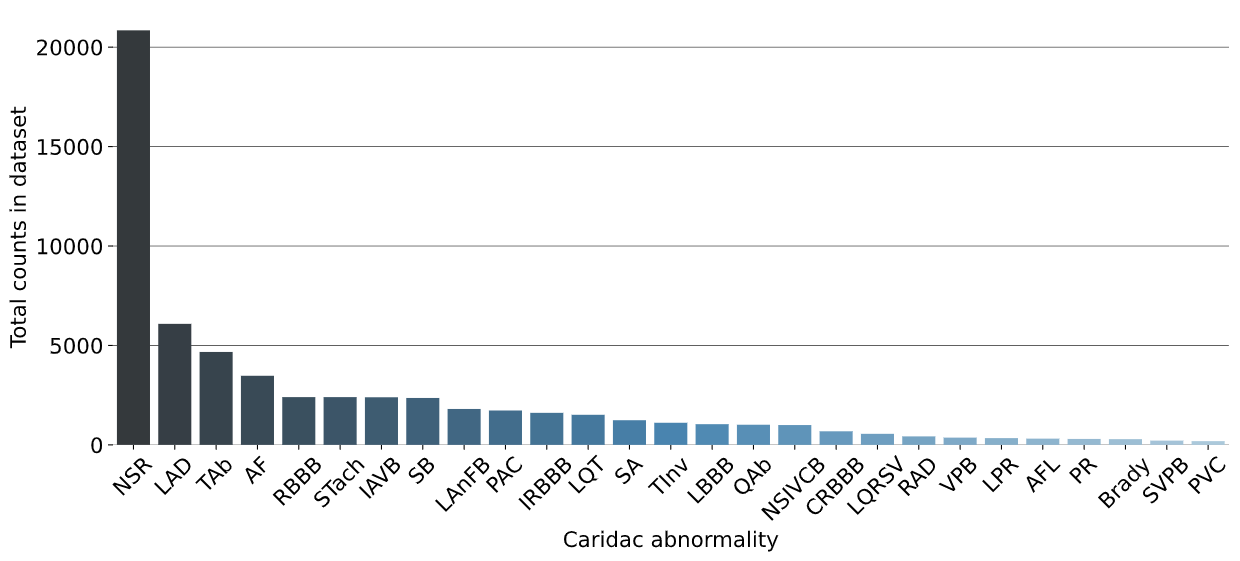}
\caption{Total counts of 27 ECG abnormalities in the original dataset, from which we observed a significant class imbalance of the dataset. Sinus rhythm (NSR) is present in more than 20,000 recordings whereas PVC were detected in less than 200 samples. Such imbalance could undermine the model’s performance, as the model is likely to learn the pattern from categories with a large number of samples while ignoring the minorities.}

\label{fig:4}
\end{centering}
\end{figure}

\textbf{Sign Loss.} A significant issue observed in our data was class imbalance, shown in Figure \ref{fig:4}, which resulted in predictions biased towards the majority class. 

Inspired by
\cite{sun2019system} , we designed an improved multi-label Sign Loss for our model training. The loss is defined as follows:

$$
\operatorname{\textit{Sign}}(p)=\left\{\begin{array}{ll}
y - 2py + p^{2} &,|y-p|<0.5 \\
1 &,|y-p| \geq 0.5
\end{array}\right.
$$
$$
\textit{Loss}=\sum_{i=1}^{27} \operatorname{\textit{Sign}}\left(p_{i}\right) \times \textit {BinaryCrossEntropyLoss}\left(p_{i}, y_{i}\right)
$$

For the correctly classified labels, a coefficient smaller than 1 was multiplied to the default binary cross-entropy loss. By doing so, the accumulated loss from a large number of true negative labels became smaller, and the loss from the misclassified labels became more prominent. Furthermore, the gradient of this loss function has a significant change around 0.5, which enables our models to capture this change, and thus the optimal binarization threshold will also be close to 0.5 and more robust.

\subsection{Evaluation Metrics}

We followed the official evaluation metrics from PhysioNet/Computing in Cardiology Challenge 2020 \cite{PhysioNet2020}. There are 111 kinds of abnormal ECG labels.  To conform the real-world clinical practice, where some misdiagnoses are less harmful than others, the misdiagnoses that end in similar outcomes or treatments as the ground truth diagnoses will still be awards with partial credit. Only 27 anomalies of the total 111 anomalies in 6 datasets were included in the final evaluation.
To be more specific, $C=\left[c_{i}\right]$ defined as the collection of our predictions. The multiclass confusion matrix is $ A=\left[a_{i j}\right] $, where $a_{i j}$ is the normalized number of recordings in a database that were classified as belonging to class $c_{i}$ but actually belong to class $c_{j}$, where $c_{i}$ and $c_{j}$ can be the same class or different classes. A reward metrics $W=\left[w_{i j}\right]$ is defined, where $w_{i j}$ denotes the reward for a positive classifier output for class $c_{i}$ with a positive label $c_{j}$. The unnormalized score will be calculated by equation \ref{eq:1}.

\begin{equation}\label{eq:1}
Unnormalized\_Score=\sum_{i=1}^{m} \sum_{j=1}^{m} w_{i j} a_{i j}
\end{equation}

\begin{equation}\label{eq:2}
Normalized\_Score=\frac{Unnormalized\_Score - Inactive\_Score}{Correct\_Score - Inactive\_Score} \\
\end{equation}

After normalization by \ref{eq:2}, a score of 1 will be assigned to the classifier that always predict the true label, while a score of 0 will be assigned to inactive classifier. The Inactive Score is the score for an inactive classifier that always outputs a normal class, while Correct Score is the score for the model that always predicts the true class. The detailed calculation of Normalized\_Score shown in \cite{eval2020}. The Normalized Score will be in range between 0 and 1, and the higher the score indicates the better performance of the model. We will evaluate the Normalized Score in our offline validation set and Hefei validation set.

\subsection{Training Setup}

The ensembled model was trained for 19 epochs with a batch size of 16 on a machine with 117 GB RAM, 4-core CPU, and one NVIDIA V100 GPU. The model parameters were optimized with the Adam optimizer \cite{adam}. The learning rate during training was set as 0.001, and rescheduled to 0.0001 at the 13\textsuperscript{th} epoch. The optimal binarization threshold was found to be 0.36 on the offline test set.

%% file: 3_results.tex
\subsection{Online Testing Results}
Table \ref{tab:1} shows the online evaluated challenge scores and rank on a) Official Validation Set, b) Official Test Database 1 and c) Official Test Database 2, d) Official Test Database 3 and e) The entire official test database.

\begin{table}[t]
\centering
\begin{tabular}{|l|c|c|}
\hline
\multicolumn{1}{|c|}{Online Test Dataset} & Score & Official Challenge Ranking \\ \hline
Official Validation Set & 0.682 & 3 \\ \hline
Official Test Database 1  & 0.852 & 3 \\ \hline
Official Test Database 2 & 0.649 & 2 \\ \hline
Official Test Database 3 & 0.396 & 3 \\ \hline
\textbf{The entire official test database} & \textbf{0.514} & \textbf{3} \\ \hline
\end{tabular}
\caption{\label{tab:1} Performance of our best online model on online datasets.}
\end{table}

\subsection{Offline Validation Results}

\begin{table}[t]
\centering
\begin{tabular}{|l|l|c|c|}
\hline
\multicolumn{1}{|c|}{Model} & Model description & \makecell[c]{Offline\\validation\\score} & \makecell[c]{Hefei\\validation\\ score} \\ \hline
Model1: Baseline SE\_ResNet & Baseline: SE\_ResNet & 0.682 & 0.241 \\ \hline

Model2: \makecell[c]{Model1+\\ QRST rule+\\ Preprocessing} & \makecell[l]{1. Add CPSC relabeling \\ and Pre-Processing; \\ 2. Add rule-based model \\ for bradycardia;} & 0.689 & 0.249 \\ \hline

Model3: \makecell[c]{Model2+\\ Sign Loss+\\ SNR Post-Processing} & \makecell[l]{1. Using Sign Loss;\\ 2. Using SNR Post-Processing;} & 0.673 & 0.300  \\ \hline

Model4: \makecell[c]{Model3+\\8 leads+\\15,000 length} & \makecell[l]{1. Using signal length of 15,000;\\ 2. Using 8 leads;\\ 3. Using Sign Loss;} & 0.674 & 0.300  \\ \hline

Model5: \makecell[c]{Model3+\\8 leads+\\5,000 length} & \makecell[l]{1. Using signal length of 5,000;\\ 2. Using 8 leads;\\ 3. Using Sign Loss;} & 0.674 & 0.236  \\ \hline

\textbf{Model6}: \makecell[c]{Ensemble \\ Model4 and Model5} & \makecell[l]{Ensemble model \textbf{(best model)}:\\ Ensemble Model4 and Model5} & \textbf{0.683} & \textbf{0.319}  \\ \hline

\end{tabular}
\caption{\label{tab:2} Performance of different models on offline validation datasets, all the models are developed based on our baseline model, from this table we select Model 6 as our best model.}
\end{table}

Table \ref{tab:2} shows the offline performance of different models we have tried based on our baseline model. Model 1 is our baseline model that uses SE\_ResNet as framework. In Model 2, we apply wavelet denoising, and add the relabelled CPSC data to training data, the performance of Model 2 improved in both our offline validation set and Hefei validation set, compared to our baseline model. However, we found that the problem of threshold shifting still remained. In order to stable the threshold and enhance the generalization of our model, we introduce Sign Loss to Model 3 and apply SNR Post- Processing. Though Model 3 shows an inferior performance on our offline validation set, it shows better performance on the Hefei validation set. To some extent, it can be explained that Sign Loss can improve the generalization ability of the model. In order to improve the efficiency of model training, Model 4 and Model 5 only use 8 leads signal data in 12 leads. The length of Model 4 signal input is 15,000, and the length of Model 5 signal input is 5,000. The training time of both models is less than that of Model 3, and the performance remains unchanged. Considering the training efficiency, model performance, and model generalization, Model 6 integrates Model 4 and Model 5, with a score of 0.683 on the offline validation set and a score of 0.319 on the Hefei validation set. The performance is better than model 4 and Model 5 on both our offline validation set and Hefei validation set. Meanwhile, the difference between the online validation score and the offline score is only 0.001 (online score: 0.682), which shows a strong ability of generalization. Finally, we select Model 6 as our best model.

\subsection{Detailed Model Performance Analysis}

\begin{figure}[t]
\begin{centering}
\includegraphics[width=\textwidth]{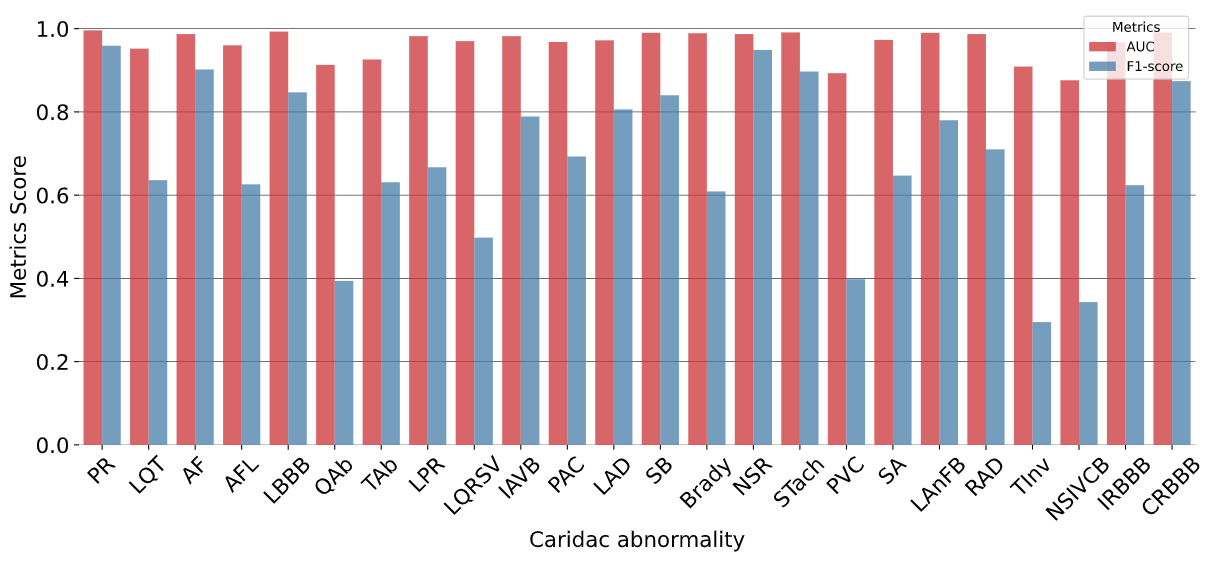}
\caption{The performance of the best model on each ECG abnormality, in terms of Area Under Curve (AUC) and F1 Score. The AUC measures the model’s ability in identifying the positive and negative samples when considering each cardiac abnormality. The F1 Score evaluates the model’s performance of multi-label classification, which takes all abnormalities into consideration. From the figure we can see that the AUC for each cardiac abnormality is relatively high, indicating that our model can classify each cardiac abnormality well when there is no intervention from other abnormalities. The fluctuation of the F1 score shows that when takes all abnormalities into account, the model’s performance decreased. This could due to different ECG abnormalities may have similar feature space, thus confuse and affect the model.}
\label{fig:5}
\end{centering}
\end{figure}

There are 27 ECG abnormalities in the original evaluation, of which 3 pairs were treated as the same ECG abnormality when calculating the score. These 3 pairs are complete right bundle branch block (CRBBB) and right bundle branch block (RBBB), premature atrial contraction (PAC) and supraventricular premature beats (SVPB), premature ventricular contradictions (PVC), and ventricular premature beats (VPB). Based on this design, we analysed our model’s performance in 24 categories. Figure \ref{fig:5} shows the performance of our proposed method on each ECG abnormality, from which we are able to find the factors that affect the model’s performance, shown below.

1) \textbf{Partial label}: The AUC of each ECG abnormality is generally high, while the f1-score of some ECG abnormality is at a low level. It is likely that some anomalies in the data are not labelled, which leads to an excessive prediction of false positive. Here are two possible reasons. Firstly, there are 6 datasets in total, each dataset has only partial abnormalities, and no dataset has all 27 abnormalities. For example, atrial fibrillation and sinus rhythm appear in all six data sets, with complete annotation and good overall model performance. However, premature ventricular contractions and low QRS voltages only appear in two data sets, thus the model’s performance is relatively poor. The second reason could be annotation error. Where some abnormalities in a dataset are missing-labelled or wrongly labelled.


2) \textbf{Hard detected features}: Some features of ECG abnormalities are hard to detect. For example, for some cases in low QRS voltages, we found that the amplitude of the signal differs, this could be due to the difference in weights and heights habitus of patients.


3) \textbf{Feature confusion}: We also found that features between two ECG abnormalities could be too similar for the model to classify. For instance, the features of bradycardia are similar to sinus bradycardia, since both of them show a slow heart rate.

As we mentioned previously, the SE module in SE\_ResNet can obtain the importance of each feature by learning, and then enhance the useful features according to the weightage and suppress the features that are not useful for the current task, so the performance   of the model can be improved compared with the original ResNet.

In our proposed method, we integrate two SE\_ResNet, which take 8-Lead ECG data data with a length of 5,000 and 15,000 as input respectively. The integrated model outperforms the two sub-models, which shows that the integrated model effectively combines the advantages of the two different input length settings. Firstly, the length of the ECG samples in the datasets ranges from 2,500 to over 100,000. The larger input length can contain more information in samples with longer signal length, while the smaller input length can make samples with short length free from the information losses of padding in training. Secondly, some of the ECG abnormalities show a characteristic of continuous repeat, while some occasionally appear. Hence, the larger input length can capture the intermittent abnormal signal, while the smaller input can reduce the difficulty of the neural network for anomaly detection.

%% file: 4_discussion.tex
The results show our ensembled approach demonstrated its ability to classify the ECG abnormalities despite the challenges presented, e.g., noise in the signals and labels. The score on the offline validation set is 0.683, only differed the online validation score by 0.001, suggesting good generalizability and little overfitting.

In this study, we also experimented with several other ideas. Two ideas that we would like to share and hopefully inspire further explorations are the segmentation of abnormal heartbeats.

\textbf{Abnormal heartbeats segmentation via 1D U-net.}
Some of the ECG abnormalities were intra-beat instead of inter-beat, and could only show up in a few heartbeats out of all. We hypothesized that given the intra-beat ECG abnormalities labels, the model could pick up them more precisely and effectively, hence resulting in a better model performance and generalization. \par
Based on the preliminary analysis of the prediction results and advice from a clinician, we selected data with premature ventricular complex (PVC) labels to annotate. The raw lead-II signals were transformed into images by plotting on a grid background. The images were then imported into Colabeler, and we manually annotated all suspected PVC heartbeats by specifying their x-axis spans. Lastly, we translated the x-axis spans back to the actual locations on the signals. Due to time constraints, only 160 PVC data were annotated and used for subsequent training. We adapted U-net \cite{unet}, a popular segmentation model in medical imaging to segment the PVC heartbeats. The 160 annotated PVC records were used as positive samples and 500 randomly selected records without PVC labels were used as negative samples to train the model.

\textbf{Feature engineering with Catboost.} Deep learning model can only automatically learn abstract features according to given tags, while feature engineering can be a supplement to abstract features. We use biosppy software package to extract r-peaks of ECG signals. Based on the extracted r-peaks, we then use the hrvanalysis software package to extract 32-dimensional features, which including 16-dimensional time-domain features and 7-dimensional frequency-domain features. Age and gender are added as features as well.

\begin{table}[t]
\centering
\begin{tabular}{|l|c|c|}
\hline
Model & Offline validation score & Hefei validation score \\ \hline
Model1: Baseline & 0.682 & 0.241 \\ \hline
\textbf{Model6}: \textbf{Best model} & \textbf{0.683} & \textbf{0.319} \\ \hline
Model7: Catboost & 0.677 & 0.249 \\ \hline
Model8: 1D U-net & 0.671 & 0.247\\ \hline
\end{tabular}
\caption{\label{tab:4} The performance of the Catboost and the 1D U-net, compared with the baseline model and the best model.}
\end{table}

Table \ref{tab:4} shows the performance of two models we have attempted, compared to our baseline model. Model 7 combines the engineered features of the ECG signal with the deep features from the neural network, and train a binary classifier for each ECG abnormality with Catboost. The offline score of the model is 0.677. The advantage of Model 7 is that it uses the engineered features of the ECG signal, which helps to classify certain ECG abnormalities more accurately. The disadvantage is that the framework of the model is complex, and the extraction of ECG signal features takes a long time. It requires too much computing resources and its time consuming, thus we have only tested this framework on our offline validation set. \par Model 8 introduces U-net to improve the classification performance for PVC.  The PVC classification is considered positive if there is any positive PVC signal output. When we incorporated U-net into our system, the PVC predictions were solely determined by U-net. Our experiments showed that incorporating U-net increased the  F\textsubscript{beta} and G\textsubscript{beta} measures in the evaluation metrics, but no challenge score improvement was observed. Due to the additional training time required, we did not incorporate U-net in our final system. \par

We plan to improve our work in the following aspects. 1) To combine ECG medical knowledge with deep learning to construct a classification model. For example, machine learning methods on R-R interval detection is not accurate. These features can be integrated with the features extracted from SE\_ResNet. Meanwhile, medical knowledge can effectively improve the interpretability and generalizability of the model. 2) To develop a better model for locating the abnormal signals. Many ECG abnormalities occur intermittently, it is a challenging task to locate the timing of abnormal signal in a long period of ECG data. An abnormality detector is desired to extract the important episodes from lengthy ECG data.

%% file: 5_conclusions.tex
In this paper, we proposed a deep learning framework to automatically identify multiple ECG abnormalities. Compared to previous works, the main contribution of our methods is three-fold. Firstly, our proposed framework is able to classify 27 types of ECG abnormalities on 12-lead ECG signals, while previous works focused on at most 9 types of ECG abnormalities classification using single lead ECG signals. Secondly, we introduce a Sign Loss to mitigate the class imbalance problem and improve the generalizability of our framework. Thirdly, our framework is developed on 6 different datasets across the various countries, and we proposed several pre-processing methods to target the diversity issue from different data sources, whereas previous works mainly use a small dataset from a single data source. Our proposed framework is developed and validated on real datasets, thus we believe it has the potential to be deployed in clinical practice.

%% file: 6_acknowledgments.tex
This work is partially supported by the National University of Singapore Start-up grant (R-608-000-172-133).

%% file: refs.bib
@article{PhysioNet2020,
  title={Classification of 12-lead ecgs: the physionet/computing in cardiology challenge 2020},
  author={Alday, Erick A Perez and Gu, Annie and Shah, Amit J and Robichaux, Chad and Wong, An-Kwok Ian and Liu, Chengyu and Liu, Feifei and Rad, Ali Bahrami and Elola, Andoni and Seyedi, Salman and others},
  journal={Physiological Measurement},
  year={2020},
  publisher={IOP Publishing}
}

@book{goldberger_goldberger_shvilkin_2018, place={Philadelphia, PA}, title={Goldbergers clinical electrocardiography: a simplified approach}, publisher={Elsevier}, author={Goldberger, Ary Louis and Goldberger, Zachary D. and Shvilkin, Alexei}, year={2018}}

@article{bickerton_pooler_2019, title={Misplaced ECG electrodes and the need for continuing training}, volume={14}, DOI={10.12968/bjca.2019.14.3.123}, number={3}, journal={British Journal of Cardiac Nursing}, author={Bickerton, Martin and Pooler, Alison}, year={2019}, pages={123–132}}

@article{kligfield2007recommendations,
  title={Recommendations for the standardization and interpretation of the electrocardiogram: part I: the electrocardiogram and its technology a scientific statement from the {A}merican {H}eart {A}ssociation Electrocardiography and Arrhythmias Committee, Council on Clinical Cardiology; the {A}merican College of Cardiology Foundation; and the {H}eart {R}hythm {S}ociety endorsed by the {I}nternational {S}ociety for {C}omputerized {E}lectrocardiology},
  author={Kligfield, Paul and Gettes, Leonard S and Bailey, James J and Childers, Rory and Deal, Barbara J and Hancock, E William and Van Herpen, Gerard and Kors, Jan A and Macfarlane, Peter and Mirvis, David M and others},
  journal={Journal of the American College of Cardiology},
  volume={49},
  number={10},
  pages={1109--1127},
  year={2007},
  publisher={Journal of the American College of Cardiology}
}

@article{kligfield2002centennial,
  title={The centennial of the {E}inthoven electrocardiogram},
  author={Kligfield, Paul},
  journal={Journal of Electrocardiology},
  volume={35},
  number={4},
  pages={123--129},
  year={2002},
  publisher={Churchill Livingstone}
}

@ARTICLE{pan32real,
  author={J. {Pan} and W. J. {Tompkins}},
  journal={IEEE Transactions on Biomedical Engineering}, 
  title={A Real-Time QRS Detection Algorithm}, 
  year={1985},
  volume={BME-32},
  number={3},
  pages={230-236},
  doi={10.1109/TBME.1985.325532}}

@article{SEResNet, title={Squeeze-and-Excitation Networks}, DOI={10.1109/cvpr.2018.00745}, journal={2018 IEEE/CVF Conference on Computer Vision and Pattern Recognition}, author={Hu, Jie and Shen, Li and Sun, Gang}, year={2018}}

@article{ResNet, title={Deep Residual Learning for Image Recognition}, DOI={10.1109/cvpr.2016.90}, journal={2016 IEEE Conference on Computer Vision and Pattern Recognition (CVPR)}, author={He, Kaiming and Zhang, Xiangyu and Ren, Shaoqing and Sun, Jian}, year={2016}}

@inproceedings{adam,
  author    = {Diederik P. Kingma and
               Jimmy Ba},
  editor    = {Yoshua Bengio and
               Yann LeCun},
  title     = {Adam: {A} Method for Stochastic Optimization},
  booktitle = {3rd International Conference on Learning Representations, {ICLR} 2015,
               San Diego, CA, USA, May 7-9, 2015, Conference Track Proceedings},
  year      = {2015},
  url       = {http://arxiv.org/abs/1412.6980},
  bibsource = {dblp computer science bibliography, https://dblp.org}
}

@misc{hefei, 
    author = {},
    year = {2020},
    title={TIANCHI - Hefei Hi-tech Cup ECG Intelligent Competition},
    url={https://tianchi.aliyun.com/competition/entrance/231754/introduction}
}

@article{unet, title={U-Net: Convolutional Networks for Biomedical Image Segmentation}, DOI={10.1007/978-3-319-24574-4_28}, journal={Lecture Notes in Computer Science Medical Image Computing and Computer-Assisted Intervention – MICCAI 2015}, author={Ronneberger, Olaf and Fischer, Philipp and Brox, Thomas}, year={2015}, pages={234–241}}

@article{1991,
  title={Detection of atrial fibrillation using artificial neural networks},
  author={Artis, Shane G and Mark, RG and Moody, GB},
  year={1991}
}

@article{VAFAIE2014291,
title = "Heart diseases prediction based on ECG signals’ classification using a genetic-fuzzy system and dynamical model of ECG signals",
journal = "Biomedical Signal Processing and Control",
volume = "14",
pages = "291 - 296",
year = "2014",
issn = "1746-8094",
doi = "https://doi.org/10.1016/j.bspc.2014.08.010",
url = "http://www.sciencedirect.com/science/article/pii/S1746809414001311",
author = "M.H. Vafaie and M. Ataei and H.R. Koofigar"
}

@article{martinez2004wavelet,
  title={A wavelet-based ECG delineator: evaluation on standard databases},
  author={Mart{\'\i}nez, Juan Pablo and Almeida, Rute and Olmos, Salvador and Rocha, Ana Paula and Laguna, Pablo},
  journal={IEEE Transactions on biomedical engineering},
  volume={51},
  number={4},
  pages={570--581},
  year={2004},
  publisher={IEEE}
}

@article{minami1999real,
  title={Real-time discrimination of ventricular tachyarrhythmia with Fourier-transform neural network},
  author={Minami, Kei-ichiro and Nakajima, Hiroshi and Toyoshima, Takeshi},
  journal={IEEE transactions on Biomedical Engineering},
  volume={46},
  number={2},
  pages={179--185},
  year={1999},
  publisher={IEEE}
}

@inproceedings{mahmoodabadi2005ecg,
  title={ECG feature extraction using Daubechies wavelets},
  author={Mahmoodabadi, SZ and Ahmadian, Alireza and Abolhasani, MD},
  booktitle={Proceedings of the fifth IASTED International conference on Visualization, Imaging and Image Processing},
  pages={343--348},
  year={2005}
}

@inproceedings{alexakis2003feature,
  title={Feature extraction and classification of electrocardiogram (ECG) signals related to hypoglycaemia},
  author={Alexakis, C and Nyongesa, HO and Saatchi, R and Harris, ND and Davies, C and Emery, C and Ireland, RH and Heller, SR},
  booktitle={Computers in Cardiology, 2003},
  pages={537--540},
  year={2003},
  organization={IEEE}
}

@article{chen2018classification,
  title={Classification of short single-lead electrocardiograms (ECGs) for atrial fibrillation detection using piecewise linear spline and XGBoost},
  author={Chen, Yao and Wang, Xiao and Jung, Yonghan and Abedi, Vida and Zand, Ramin and Bikak, Marvi and Adibuzzaman, Mohammad},
  journal={Physiological measurement},
  volume={39},
  number={10},
  pages={104006},
  year={2018},
  publisher={IOP Publishing}
}

@article{GAZIANO201072,
title = "Growing Epidemic of Coronary Heart Disease in Low- and Middle-Income Countries",
journal = "Current Problems in Cardiology",
volume = "35",
number = "2",
pages = "72 - 115",
year = "2010",
note = "Growing Epidemic of Coronary Heart Disease in Low- and Middle-Income Countries",
issn = "0146-2806",
doi = "https://doi.org/10.1016/j.cpcardiol.2009.10.002",
url = "http://www.sciencedirect.com/science/article/pii/S0146280609001273",
author = "Thomas A. Gaziano and Asaf Bitton and Shuchi Anand and Shafika Abrahams-Gessel and Adrianna Murphy"
}

@article{lecun1995convolutional,
  title={Convolutional networks for images, speech, and time series},
  author={LeCun, Yann and Bengio, Yoshua and others},
  journal={The handbook of brain theory and neural networks},
  volume={3361},
  number={10},
  pages={1995},
  year={1995}
}

@article{xiong2018ecg,
  title={ECG signal classification for the detection of cardiac arrhythmias using a convolutional recurrent neural network},
  author={Xiong, Zhaohan and Nash, Martyn P and Cheng, Elizabeth and Fedorov, Vadim V and Stiles, Martin K and Zhao, Jichao},
  journal={Physiological measurement},
  volume={39},
  number={9},
  pages={094006},
  year={2018},
  publisher={IOP Publishing}
}

@article{sodmann2018convolutional,
  title={A convolutional neural network for ECG annotation as the basis for classification of cardiac rhythms},
  author={Sodmann, Philipp and Vollmer, Marcus and Nath, Neetika and Kaderali, Lars},
  journal={Physiological measurement},
  volume={39},
  number={10},
  pages={104005},
  year={2018},
  publisher={IOP Publishing}
}

@article{warrick2018ensembling,
  title={Ensembling convolutional and long short-term memory networks for electrocardiogram arrhythmia detection},
  author={Warrick, Philip A and Homsi, Masun Nabhan},
  journal={Physiological measurement},
  volume={39},
  number={11},
  pages={114002},
  year={2018},
  publisher={IOP Publishing}
}

@misc{eval2020, 
    author = {},
    year = {2020},
    title={PhysioNet/Computing in Cardiology Challenge 2020 Evaluation},
    url={https://github.com/physionetchallenges/evaluation-2020/blob/master/evaluate\_12ECG\_score.py}
}

@inproceedings{0Comparing,
  title={Comparing feature-based classifiers and convolutional neural networks to detect arrhythmia from short segments of ECG},
  author={Andreotti, Fernando and Carr, Oliver and Pimentel, Marco AF and Mahdi, Adam and De Vos, Maarten},
  booktitle={2017 Computing in Cardiology (CinC)},
  pages={1--4},
  year={2017},
  organization={IEEE}
}

@inproceedings{0Detection,
  title={Detection of AF and other rhythms using RR variability and ECG spectral measures},
  author={Billeci, Lucia and Chiarugi, Franco and Costi, Magda and Lombardi, David and Varanini, Maurizio},
  booktitle={2017 Computing in Cardiology (CinC)},
  pages={1--4},
  year={2017},
  organization={IEEE}
}

@inproceedings{2017Detection,
  title={Detection of atrial fibrillation using decision tree ensemble},
  author={Bin, Guangyu and Shao, Minggang and Bin, Guanghong and Huang, Jiao and Zheng, Dingchang and Wu, Shuicai},
  booktitle={2017 Computing in Cardiology (CinC)},
  pages={1--4},
  year={2017},
  organization={IEEE}
}

@incollection{2019Arrhythmia,
  title={Arrhythmia Classification with Attention-Based Res-BiLSTM-Net},
  author={Huang, Chengbin and Zhao, Renjie and Chen, Weiting and Li, Huazheng},
  booktitle={Machine Learning and Medical Engineering for Cardiovascular Health and Intravascular Imaging and Computer Assisted Stenting},
  pages={3--10},
  year={2019},
  publisher={Springer}
}

@incollection{2019Automatic,
  title={Automatic Multi-label Classification in 12-Lead ECGs Using Neural Networks and Characteristic Points},
  author={Xia, Zhourui and Sang, Zhenhua and Guo, Yutong and Ji, Weijie and Han, Chenguang and Chen, Yanlin and Yang, Sifan and Meng, Long},
  booktitle={Machine Learning and Medical Engineering for Cardiovascular Health and Intravascular Imaging and Computer Assisted Stenting},
  pages={80--87},
  year={2019},
  publisher={Springer}
}

@incollection{2019A,
  title={A 12-Lead ECG Arrhythmia Classification Method Based on 1D Densely Connected CNN},
  author={Wang, Chunli and Yang, Shan and Tang, Xun and Li, Bin},
  booktitle={Machine Learning and Medical Engineering for Cardiovascular Health and Intravascular Imaging and Computer Assisted Stenting},
  pages={72--79},
  year={2019},
  publisher={Springer}
}

@incollection{0Multi,
  title={Multi-label Classification of Abnormalities in 12-Lead ECG Using 1D CNN and LSTM},
  author={Luo, Chengsi and Jiang, Hongxiu and Li, Quanchi and Rao, Nini},
  booktitle={Machine Learning and Medical Engineering for Cardiovascular Health and Intravascular Imaging and Computer Assisted Stenting},
  pages={55--63},
  year={2019},
  publisher={Springer}
}

@inproceedings{sun2019system,
  title={System-level hardware failure prediction using deep learning},
  author={Sun, Xiaoyi and Chakrabarty, Krishnendu and Huang, Ruirui and Chen, Yiquan and Zhao, Bing and Cao, Hai and Han, Yinhe and Liang, Xiaoyao and Jiang, Li},
  booktitle={2019 56th ACM/IEEE Design Automation Conference (DAC)},
  pages={1--6},
  year={2019},
  organization={IEEE}
}

@article{devlin2018bert,
  title={Bert: Pre-training of deep bidirectional transformers for language understanding},
  author={Devlin, Jacob and Chang, Ming-Wei and Lee, Kenton and Toutanova, Kristina},
  journal={arXiv preprint arXiv:1810.04805},
  year={2018}
}

@article{krizhevsky2017imagenet,
  title={Imagenet classification with deep convolutional neural networks},
  author={Krizhevsky, Alex and Sutskever, Ilya and Hinton, Geoffrey E},
  journal={Communications of the ACM},
  volume={60},
  number={6},
  pages={84--90},
  year={2017},
  publisher={ACM New York, NY, USA}
}
